\documentclass[manuscript]{aastex}

\shorttitle{Saturation of Magnetorotational Instability}
\shortauthors{Ebrahimi et al.}

\begin{document}

\title{Saturation of Magnetorotational Instability through Magnetic Field Generation}

\author{F. Ebrahimi, S. C. Prager and D. D. Schnack}
\affil{University of Wisconsin-Madison, Madison, Wisconsin, 
and Center for Magnetic-Self Organization in Laboratory and Astrophysical Plasmas}

\begin{abstract}
The saturation mechanism of Magneto-Rotational Instability (MRI) is examined
through analytical quasilinear theory 
and through nonlinear computation of a single mode in a rotating disk. 
We find that large-scale magnetic field is generated through the
alpha effect (the correlated product of velocity and magnetic field fluctuations) and 
causes the MRI mode to saturate. If the large-scale plasma flow is allowed to evolve, the mode
can also saturate through its flow relaxation. In astrophysical plasmas, for which the flow cannot
relax because of gravitational constraints, the mode saturates through field generation only. 
\end{abstract}

\section{Introduction}
\label{sec:intro}

It is well known that collisional hydrodynamic viscosity is too small to explain the inferred rate of angular momentum transport in accretion disks (Shakura \& Sunyaev 1973). One possible source of anomalous viscosity is turbulence resulting from the magneto-rotational instability, or MRI (Velikhov 1959; Balbus \& Hawley 1991). In the linear phase, this instability derives its source of free energy from the radial gradient of the rotational flow in the presence of a \textit{weak} magnetic field, i.e., the MRI requires the presence of a magnetic field, but is linearly stable if the field is too large. The properties of MRI turbulence, and its effect on angular momentum transport, depend in a fundamental way on the amplitude of the fluctuations in the non-linear state.  We expect this amplitude to be related to the non-linear saturation mechanism of a \textit{single} MRI mode. 

A common saturation mechanism for linear instabilities is the relaxation of the free energy source through quasi-linear processes.  For the case of the MRI, this would imply modification of the radial profile of the rotational velocity, but this mechanism may not be available in accretion disks and other astrophysical settings because of strong gravitational constraints that maintain a Keplerian profile ($V_\phi \propto r^{-1/2}$).  Instead, the MRI may saturate (cease linear growth) if the non-linear evolution of the instability generates a \textit{mean} component of the magnetic field that is sufficient to stabilize the mode.

We remark that for most plasma instabilities, if the energy source is fixed and unvarying (and nonlinear coupling to other modes is ignored), the mode will grow without bound, until limited by dissipation or some physical dimension of the system. In this sense, the MRI might be atypical.  

In this paper we investigate the linear growth and non-linear saturation mechanism and amplitude of a \textit{single} MRI mode in thick-disk geometry, i.e., $\frac{L}{r} \sim 1$.  Our primary tool is linear and non-linear MHD computation in cylindrical ($r, \phi , z$) geometry which is periodic in the azimuthal $(\phi)$ and axial ($z$) directiona.  We solve a model initial value problem in which the inner and outer radial boundaries are impermeable, perfectly conducting, concentric cylinders that can rotate independently at specified rates.  They are coupled to the internal flow by hydrodynamic viscosity. The initial mean (azimuthally and axially averaged) profile $V_\phi (r)$ is Keplerian, and in most cases is assumed to be maintained for all times by external forces.  Perturbations to the equilibrium that depend on ($r, \phi , z$)  are then  introduced, and evolve dynamically according to the single fluid MHD equations.  Linear growth rates and eigenfunctions are found by integrating the linearized, single fluid, visco-resistive MHD equations forward in time until an exponentially growing solution is obtained.  The saturation mechanism and amplitude of the mode are determined by solving the non-linear MHD equations for a single mode, with azimuthal mode number $m$ and axial mode number $k$, beginning from a small initial perturbation.  At finite amplitude, the mode will interact with itself to modify the mean background state, which in turn alters both its temporal evolution and its radial dependence.  In this sense it differs from a quasi-linear calculation, in which only the effect of the \textit{linear} eigenfunction on the background is accounted for.  

Our \textit{primary result} is that, when the \textit{mean} flow profile is maintained as Keplerian, the MRI can saturate due to the generation of a mean component of the magnetic field that is sufficiently large to linearly stabilize the instability.  
The mechanism for field generation is by means of a \textit{mean} 
electric field 
$<\textbf{E}>$ = $-< \widetilde{\textbf{V}} \times  \widetilde{\textbf{B}}  >$ =$\alpha <\textbf{B}>$ 
(where $< \cdots >$ denotes the azimuthal and axial average; and $\widetilde{\textbf{V}}$ and $\widetilde{\textbf{B}}$ are the velocity and magnetic fluctuations) that is produced by the nonlinear interaction of the finite amplitude mode with itself, and is parallel to the mean magnetic field. This is the well known $\alpha$-effect, the correlated product of velocity and magnetic field fluctuations (Moffatt 1978). It is a fundamental dynamo mechanism. This suggests that the MRI may operate as a dynamo for the generation magnetic fields in accretion disks and other rotating systems.  This is consistent with earlier results indicating that the magnetic energy increases during the onset of MRI turbulence (Hawley, Gammie \& Balbus 1996).   

Finally, when the mean flow profile is allowed to evolve, a single MRI mode saturates by relaxation of the rotation profile, as expected. This is also accompanied by the generation of a mean magnetic field.  Thus, the MRI dynamo mechanism (the alpha effect) may be robust.
Our computational results are supported by an analytic theory in which the quasi-linear stresses and mean parallel electric field are calculated explicitly in terms of the linear eigenfunctions of the MRI.

The paper is organized as follows. Previous work and the model 
are described in sections \ref{sec:pwrk} and \ref{sec:model} respectively. 
Nonlinear evolution of a single axisymmetric MRI mode is presented in section \ref{sec:nl},
for fixed Keplerian flow (section \ref{sec:nl1}) and for evolving flow (section \ref{sec:nl2} ).
 Using nonlinear single mode computations, 
the alpha effect for non-axisymmetric MRI modes is examined in 
section \ref{sec:asynl}. 
In section \ref{sec:ql}, 
the alpha effect for axisymmetric and non-axisymmetric MRI 
is obtained using quasilinear calculations.
We summarize in section \ref{sec:summ}.

\section{Previous Work}  
\label{sec:pwrk}

In early work, the linear theory of the MRI has been 
investigated carefully, and in the past decade nonlinear MHD 
simulations of the MRI have been performed (see, e. g., 
Balbus \& Hawley 1998; Hawley \& Balbus 1999 and the references
therein), including 
evolution of an MRI-unstable system to a turbulent state. 
An interesting feature of the turbulent state is that 
small-scale magnetic fluctuations arise, representing an increase 
in magnetic energy (Hawley, Gammie \& Balbus 1996). 
Turbulence generated by MRI in global disk simulations
is one of the current MRI research topics (Armitage 1998; 
Machida, Hayashi \& Matsumoto 2000; Hawley 2000; Hawley 2001;
 Machida \& Matsumoto 2003; Reynolds \& Armitage 2001; Armitage \& Reynolds 2003; McKinney \& Gammie 2002; Nishikori, Machida \& Matsumoto 2006; Fromang \& Nelson 2006; Lyra et al. 2008).

Recently the issue of numerical convergence has been raised in the simulations of the MRI. 
It was found that the results of some simulations are not independent
of the numerical dissipation scale, and physical dissipations should be 
used (Fromang \& Papaloizou 2007).  
Numerical simulations of the MRI-induced turbulence in a 
shearing box with nonzero net vertical flux (Lesure \& Longaretti 2007) 
and zero net flux (Fromang, Papaloizou, Lesur \& Heinemann 2007) showed that 
the physical dissipation, the viscosity and the resistivity, affect the 
saturated level of MRI-driven 
MHD turbulence, and thus the amount of angular momentum
transport. It was found that the turbulent activity is an increasing function of the 
magnetic Prandtl number. The dependence of the saturation level of the MRI
on gas pressure has been investigated by Sano et al. 2004 
using the local approximation. Based on series of local numerical 
simulations, a scaling law was also provided that describes the transport of angular 
momentum in shearing MHD boxes (Pessah, Chan \& Psaltis 2007).

In addition to numerical simulations, analytical methods were used to study some nonlinear aspects of the MRI by Goodman \& Xu 1994. Goodman \& Xu introduced the notion of whether the MRI can amplify very weak initial fields until they reach equipartition with the gas pressure. They showed that MRI axisymmetric modes are exact nonlinear solutions of the local incompressible MHD equations, suggesting that the MRI perturbed magnetic field may grow to an amplitude many times larger than the initial unperturbed field. Their analytical calculations assumed a very special form for the axisymmetric MRI mode, with a purely vertical wavevector whose field perturbations are purely horizontal (perpendicular to the rotation axis). In this paper, we solve the full nonlinear MHD equations for a single MRI mode with perturbations in all three directions (radial, azimuthal and vertical).  We find that large-scale mean magnetic field is amplified through the alpha effect that causes the mode to saturate. This effect was not studied by Goodman \& Xu. In fact, the assumption made by Goodman \& Xu may have limited the possibility of the MRI mode saturation through the alpha effect, which is the central result of our paper.  In the limit of large amplification of MRI perturbation, Goodman \& Xu also showed that the MRI mode could themselves be subject to additional parasitic instabilities (such as the Kelvin-Helmholtz instability) which may destroy the primary MRI modes. The growth rate of parasitic instabilities is proportional to the MRI perturbation amplitude. Simulations of these instabilities would require multiple wavenumbers, and cannot be captured in single mode (single $k_z$) computations. 

However, the saturation \textit{mechanism}, a 
fundamental feature of MRI, is not fully understood. Most of the previous studies on the MRI saturation have dealt with what controls the saturated level of the MRI in a turbulent state. The most recent results show the effect of physical 
dissipation parameters on the saturation level using an unstratified shearing box model. 
However, these studies have not fundamentally investigated the cause of the mode saturation
(Lesure \& Longaretti 2007; Fromang, Papaloizou, Lesur \& Heinemann 2007; 
Pessah, Chan \& Psaltis 2007; Sano et al. 2004). 
 In this paper, we identify and explain the saturation mechanism of the MRI 
at the level of a single mode. Through the nonlinear evolution 
of a single mode, we simplify the problem to better identify the root of 
MRI saturation. Using this simple dynamical model, 
we find that the instability itself generates a large-scale magnetic field 
(mean magnetic field) through the alpha effect that causes the mode to saturate.

\section{The model}
\label{sec:model}
We employ the MHD equations in doubly periodic $(r,\phi,z)$ cylindrical geometry for both the analytic
 (quasi-linear)
and computational studies.  All variables are decomposed as $f(r,\phi,z,t)=\sum_{(m,k)} \widehat f_{m,k}(r,t) e^{i(m \phi + kz)}=<f(r,t)>+\widetilde f(r,\phi\,z,t)$, where $<f>$ is the \textit {mean} $(m=k=0)$ component, and $\widetilde f$ is the fluctuating component (i.e., all other Fourier components with $m \ne 0$ and $k \ne 0$).  Note that the mean component is a function of radius.  We consider an azimuthal equilibrium flow 
$\textbf{V}_0  = V_\phi (r) \hat \phi$ 
and 
vertical magnetic field  $\textbf B = B_0\hat{z}$. (Here, $V_{\phi}(r)$ is the equilibrium value of the mean azimuthal flow $< V_{\phi}>$, and $B_0$ is a constant.) The single fluid MHD equations are,  
\begin{eqnarray}
\frac {\partial \textbf A }{ \partial t } &=& -\textbf{E} = 
S\textbf V\times \textbf B - \eta \textbf J\\
\rho \frac {\partial \textbf V }{ \partial t } &=&
-S \rho \textbf V . \nabla\textbf V + S\textbf J \times
\textbf B +P_m \nabla^2 \textbf V -S \frac{\beta_0}{2}\nabla P -S \rho \nabla \Phi + \textbf F_g \\
\frac {\partial P }{ \partial t } &=&
-S\nabla \cdot (P \textbf V) - S (\Gamma -1) P \nabla \cdot \textbf V\\
\frac {\partial \rho }{ \partial t } &=&
-S\nabla \cdot (\rho \textbf V) \\
\textbf B &=& \nabla \times \textbf A\\
\textbf J &=& \nabla \times \textbf B
\end{eqnarray}
where the variables, $\rho, P, V, B, J, \Gamma$, and $\Phi$ are 
the density, pressure, velocity, magnetic field, current, ratio of the specific heats, and 
gravitational potential, respectively.
Time and radius are normalized to the resistive diffusion time 
$\tau_R = 4 \pi {r_2}^2/c^2\eta_0$ and the outer radius $r_2$, 
making the normalized outer radius unity. 
Velocity \textbf{V} and 
magnetic field \textbf{B} are normalized to the Alfv\'en velocity
 $V_A$, and the magnetic field on axis $B_0$, respectively.
The parameter $S = \frac {\tau_R}{\tau_A}$ 
is the Lundquist
number (where $\tau_A = r_2/V_A$), and $P_m= \frac {\tau_R}{\tau_{vis}}$  
 measures the ratio of 
characteristic viscosity $\nu_0$ to resistivity $\eta_0$ (the magnetic Prandtl number), 
where $\tau_{vis}= 4 \pi {r_2}^2/c^2\nu_0$
is the  viscous 
diffusion time.
The factor $\beta_0 =8\pi P_0/B_0^2$ is the beta normalized 
to the axis value. The resistivity 
profile $\eta$ is uniform.  The boundary conditions in the radial direction are as is appropriate to dissipative MHD with a perfectly conducting boundary: the tangential electric field, the normal component of the magnetic field, and the normal component of the velocity vanish, and the tangential component of the velocity is the rotational velocity of the wall.  The azimuthal $(\phi)$ and axial $(z)$ directions are periodic.

We pose an initial value problem that consists of the equilibrium plus a perturbation of the form 
$\tilde f(r, \phi,z,t)=\tilde f_{m,k} (r,t) \mathrm exp( im\phi +ikz)$. 
Equations (1-6) are then integrated forward in time using the DEBS code.   The DEBS code uses a 
finite difference method with a staggered grid for radial discretization and pseudospectral method for 
azimuthal and vertical coordinates.  In this decomposition, each mode satisfies a separate equation of the form $\partial \tilde f_{m,k} / \partial t = L_{m,k}  \tilde f_{m,k} +\sum_{(m^{'},k^{'})} N_{m,k,m^{'},k^{'}}$, where $L_{m,k}$ is a linear operator that depends on $\tilde f_{0,0}(r,t)$, and $N_{m,k,m^{'},k^{'}}$ is a nonlinear term that represents the coupling of the mode $(m,k)$ to all other modes $(m^{'},k^{'})$.  (This latter term is evaluated pseudospectrally.) The time advance is a combination of the leapfrog and semi-implicit methods (Schnack et al. 1987).    

We distinguish between three types of initial value computations.  In \textit{linear} computations, the initial conditions consist of an equilibrium $<f(r)>$ plus a single mode $\tilde f_{m,k}(r,0) \mathrm exp( im\phi +ikz)$, as described above, with $\left| \tilde f_{m,k} /<f> \right| << 1$.  The initial amplitude $\tilde f_{m,k}(r,0)$ is a polynomial in $r$ that satisfies the boundary conditions at $r=r_1$ and $r=r_2$.  The initial amplitudes of all other modes are set to zero, and are constrained to remain zero for all time. Only the mode $(m,k)$ is then evolved; in particular,  the equilibrium (the $m=0$, $k=0$ mode, or $<f>$) is \textit{not} evolved, and remains fixed in time.  The mode $(m,k)$ then satisfies the linear equation  $\partial \tilde f_{m,k} / \partial t = L_{m,k}  \tilde f_{m,k}$, where $L_{m,k}(r)$ is independent of time; an exponentially growing solution indicates linear instability, and the solution $\tilde f_{m,k}(r,t)$ is the radial profile of the unstable eigenfunction.

In \textit{nonlinear single mode} computations, the problem is initialized in the same manner.  However,  the $m=0$, $k=0$ component (the background) is allowed to evolve self-consistently.  The term $N_{0,0,m,k}$ contains the nonlinear contribution from the mode $(m,k)$ interacting with itself, and  the operator $L_{m,n}$, which depends on $\tilde f_{0,0}$, can evolve with time.  All other modes $(m^{'},k^{'})$ have and retain zero amplitude.  When the mode amplitude is small ($\left| \tilde f_{m,k} /<f> \right| << 1$), the nonlinear term is negligible and the linear solution is again obtained.  As the unstable mode grows and $\left| \tilde f_{m,k} /<f> \right|$ becomes finite, the evolution of the background profile $\tilde f_{0,0}$ can modify the term $L_{m,k}(r,t)$ and this can affect the evolution the mode $(m,k)$.  If, as a result, the mode $(m,k)$ ceases to grow exponentially it is said to \textit{saturate}.  (In most cases, this means that the mode ceases continual growth altogether.)  We remark that this differs from \textit{quasi-linear} saturation, because in the saturated state the radial profile of the finite amplitude solution $\tilde f_{m,k}(r,t)$ can differ significantly from that of the linearly unstable eigenfunction. 

In a \textit{fully nonlinear} computation, all modes are intitialized with small random amplitude and are  evolved in time, including the full nonlinear term $N_{m,k,m^{'},k^{'}}$.  In this case, saturation can occur as a result of both background modification and coupling to other modes. If there is sufficient numerical resolution, this can lead to a turbulent state.

In this work we consider both linear and nonlinear single mode computations of the MRI.  The latter begin from a state in which the plasma rotates azimuthally with a mean 
Keplerian flow $<V_{\phi}(r)> \propto r^{-1/2} $.  The initial (i.e., at $t=0$)  radial \textit {equilibrium} 
force balance (Eq. 2) is satisfied by $ \frac{\beta_0}{2}\nabla p + \rho \nabla \Phi = \rho V_{\phi}^2/r$, where $\nabla \Phi =GM/r^2$, and 
the initial pressure and density profiles are assumed to be radially uniform.
 
The azimuthal flow profile is the source of free energy for the MRI. We might expect that the MRI would saturate by altering this profile.  However, in astrophysical settings the Keplerian profile is fixed by gravitational constraints, so this saturation mechanism may not be available.  We investigate the saturation process in this case by performing  computations in which the \textit{mean} flow profile is fixed in time (see section \ref{sec:nl1}).  In these computations, the mean Keplerian profile
is maintained in time by an external force;
an ad-hoc force ($\textbf F_g$) is added in the momentum equation to
fix the mean flow profile.   In other computations we allow the mean flow profile to evolve in time (see section \ref{sec:nl2}); 
the initial Keplerian flow consistent with the angular velocity at the inner and outer 
boundaries is allowed to relax self-consistently as the mode reaches finite amplitude.  This is the usual case of relaxation of the source of free energy for the mode.  In both cases, all other mean variables (e.g., the mean magnetic field profile) are allowed to evolve.
Pressure and density are evolved; however, they remain fairly uniform during the 
computations. (Note that \textit {perturbations} to the azimuthal velocity $\widetilde V_{\phi}$ are allowed to evolve.)   

In all cases the MRI is found to cease exponential growth, and the background profile $<f(r)>$ is modified.  The saturation mechanism for a single MRI mode is then investigated by performing \textit{linear} computations in which the initial background state $f_{0,0}(r,0)$ is the saturated background profile obtained from the nonlinear single mode calculations, $<f(r)>$.  We find that this new linear solution does \textit{not} grow exponentially, indicating that, as a result of the nonlinear single mode evolution, the background state has become linearly stable.  In particular, this state has a modified mean magnetic field profile that has sufficient magnitude to stabilize the mode.

Fully nonlinear calculations of MRI saturation will be reported in a future publication.

We consider a cylindrical disk-shaped plasma with aspect ratio 
 L/($r_2 -r_1$) 
(where L is the vertical height, and $r_1$ and $r_2$ 
are the inner and outer radii, respectively). 
The inner and outer radial boundaries are perfectly conducting, concentric 
cylinders that can rotate independently at specified rates.
Periodic boundary conditions are used 
in the vertical and azimuthal directions.

The aspect ratio (L/($r_2 -r_1$))
used in the nonlinear computations is 1.3. The nonlinear computations are performed
in a thick-disk approximation where vertical and radial 
distances are of the same order.
 The range of parameters used in the computations 
is $\beta= 10^4-10^5$, $S=10^2-10^5$, $P_m=0.1-1$.

\section{Axisymmetric single mode nonlinear evolution}
\label{sec:nl}
We present the results of single mode nonlinear computations with both fixed and evolving mean azimuthal velocity profiles for an axisymmetric MRI mode (i.e., $m=0$), as these modes have the largest linear growth rate (see, e.g., Balbus \& Hawley 1998).  The fixed mean flow profile is relevant to the astrophysical setting; evolving mean flow is the usual case of relaxation of the free energy source. In both cases the initial condition consists of the equilibrium plus a perturbation with single vales of $m$ (in this case $m=0$) and $k$. In the first, we maintain 
a fixed \textit {mean} profile $<V_{\phi}>$, so as to isolate the effect of
magnetic field growth on mode saturation. In the second,  the \textit {mean} azimuthal flow profile
is permitted to evolve in time.  (In both cases, \textit {perturbations} to the azimuthal flow $\widetilde V_{\phi}$ are permitted and evolve.) 
We find that mean magnetic field growth plays an important role in the  saturation of the mode,
although the effect is somewhat stronger when the \textit {mean} azimuthal flow is fixed in time.

\subsection{Evolution with fixed mean flow}
\label{sec:nl1}
We computationally solve the MHD equations (section \ref{sec:model}) with initial conditions of
 Keplerian flow ($V_{\phi} = V_0 r^{-1/2}$) and 
a weak vertical magnetic field $\textbf B= B_0 \hat z$.
We first, consider only axisymmetric modes ($m=0$),
which are the fastest growing modes and will be treated analytically in section \ref{sec:ql}.
Since the computations are single mode (single $m$ and $k$), 
the spatial resolution is only 
in the radial direction. The number of radial grid points used is 250.
The solutions
are converged in timestep and radial spatial resolution.
We have performed nonlinear computations with different axial wave number and 
 different Lundquist numbers, $S=10^2,10^3,10^4$, and 
with different equilibrium azimuthal flow magnitudes $ V_0=10 - 100$ $V_A$.
We find that the dynamics of mode 
saturation (and field amplification) is independent of axial wave number $k$ and flow amplitudes,
 and very similar for the three different Lundquist numbers. 
Therefore, we only present nonlinear
computation for the axisymmetric $k=6$ mode with $V_0=16 V_A$, at $S=10^2$ and $S=10^3$.

\begin{figure}
\includegraphics[]{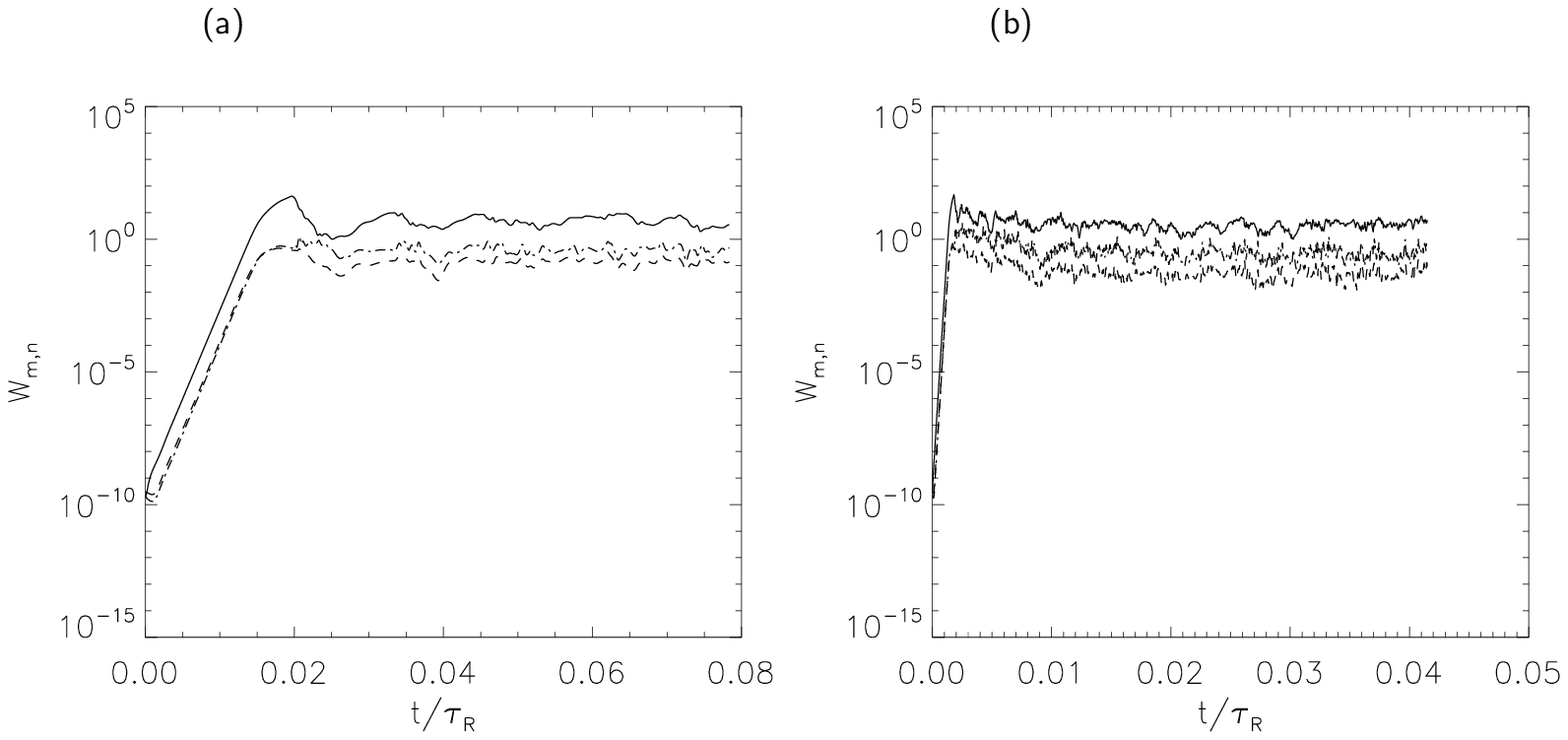}
\caption{Nonlinear evolution of axisymmetric MRI mode. 
Magnetic energies, $W_{\phi}=1/2 \int{\widetilde B_{{\phi}(m,n})^2 dr^3}$ (solid line), 
  $W_r=1/2 \int{\widetilde B_{r(m,n})^2 dr^3}$ (dashed line), and 
$W_z=1/2 \int{\widetilde B_{z(m,n})^2 dr^3}$ (dashed-dotted line) vs. time for
computation with mean flow profile fixed ($\beta =10^4, P_m=1$). (a) $S=10^2$ (b) $S=10^3$ }
\label{fig:fig1}
\end{figure}

\begin{figure}
\includegraphics[]{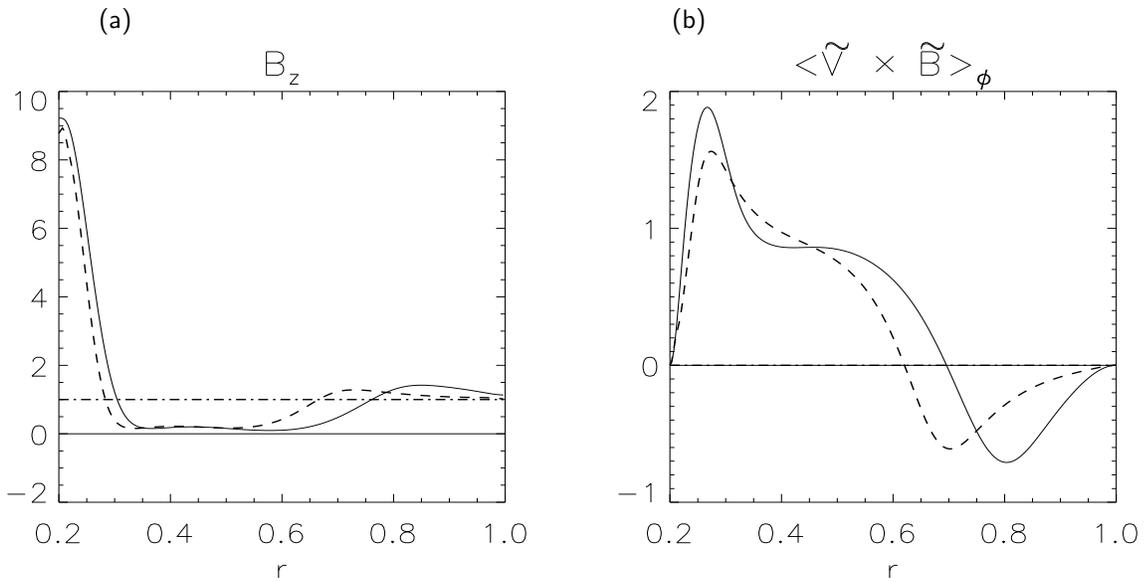} 
\caption{Radial profiles when mode starts to saturate in computation with fixed mean flow 
of (a) mean vertical magnetic field  (b) azimuthal component of the alpha effect, for $S=10^2$ (solid line) at t=0.016 and $S=10^3$ (dashed line) at t=0.0015 .}
\label{fig:fig2}
\end{figure}

Figure \ref{fig:fig1} shows the components of the magnetic modal energy 
during the nonlinear evolution when flow is fixed for $S=10^2$ and $S=10^3$.
The axisymmetric mode grows with the growth 
rate $\gamma \tau_A=9.5$ and starts to saturate around t=0.016 (Fig.~\ref{fig:fig1}(a)).
Kinetic energy at t=0 is about two order of magnitude (50 times) 
larger than magnetic energy; however after saturation of the mode, total magnetic
energy increases, and it becomes of the same order as the kinetic energy.
 Mean vertical magnetic field at t=0 is $B_z=1$, and is shown with the dashed-dotted line in
Fig.~\ref{fig:fig2}(a). 
As the mode starts to saturate, the vertical magnetic field starts to increase as the 
result of the alpha effect $<\widetilde V \times \widetilde B>_{\phi}$.
 Figure \ref{fig:fig2}(a) shows the mean vertical magnetic field
 at time t=0.016 for $S=10^2$ (around
the time of mode saturation). 
At this time the vertical mean magnetic field in the core
(r=0.2-0.4) is increased by a factor of nine. The azimuthal component of the alpha effect 
which causes the generation of
 the vertical magnetic field is large in the core (Fig. \ref{fig:fig2}(b)).
The mean vertical magnetic field and the azimuthal component of the alpha effect 
are also shown in Fig.~\ref{fig:fig2} for $S=10^3$ (dashed line) around
the time of mode saturation.
Similar radial structures and magnetic field amplification are observed for $S=10^3$.
The radial structure of the nonlinear alpha effect 
(Fig. \ref{fig:fig2}(b)) is consistent with the quasilinear calculations (section 6, Fig. \ref{fig:fig1})
and results in the amplification of magnetic field. 
Therefore, the nonlinear computation shows that the fluctuation-induced 
alpha effect amplifies the mean magnetic field in the core and causes the saturation of the mode.

\begin{figure}
\includegraphics[]{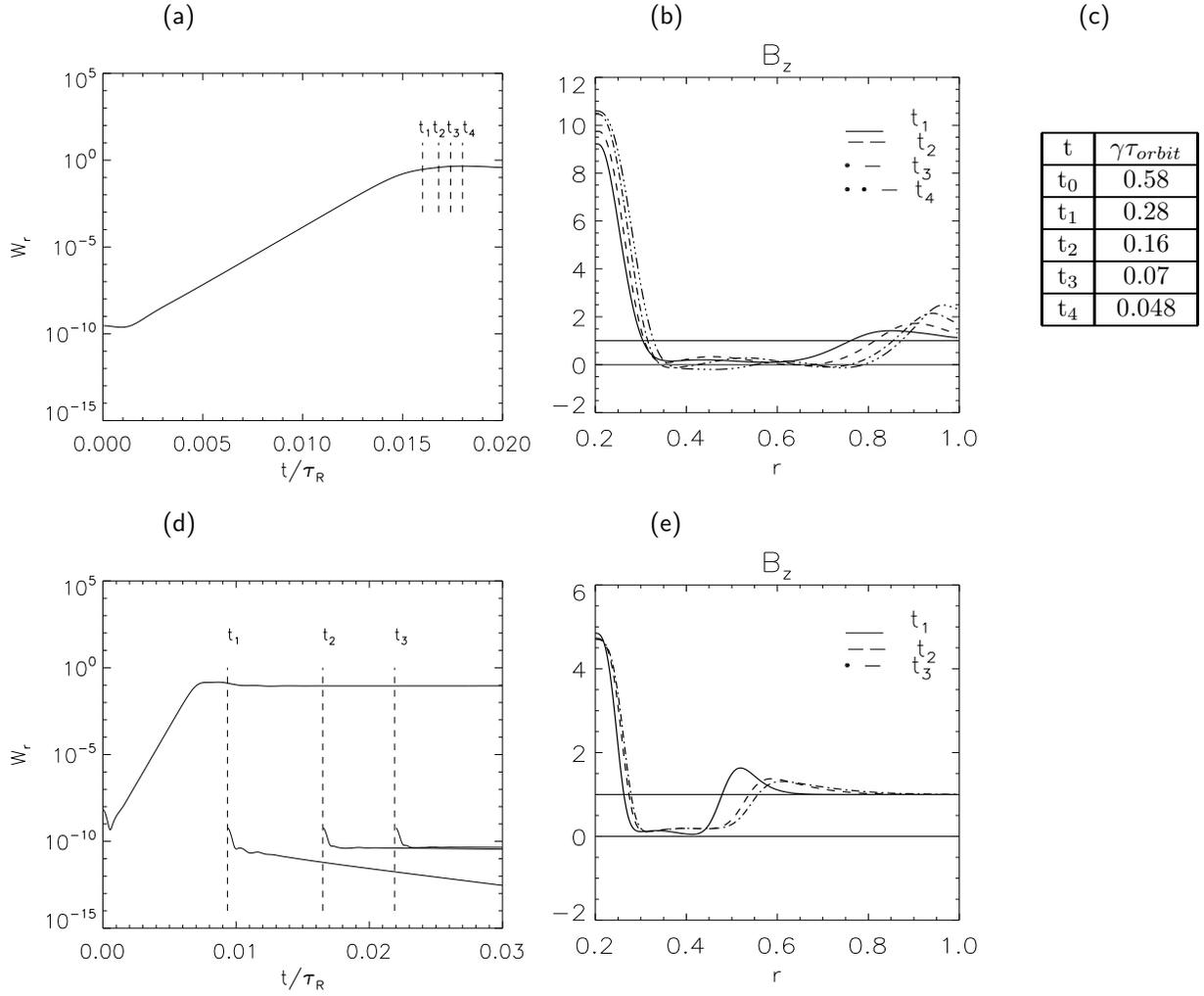} 
\caption{(a)--(c) Computations for m=0, k=6 mode. 
(d)--(e) Computations for m=0, k=30.  (a) Radial magnetic energy during the nonlinear evolution of MRI mode (m=0, k=6) with
fixed mean flow. Four times during the saturation are marked.
(b) Mean vertical magnetic field profiles for these four times during the nonlinear saturation. Four linear computations are initialized with these mean magnetic field profiles. 
 (c) Growth rates (
$\gamma \tau_{orbit}$) calculated from the linear m=0, k=6 computations. (d) 
Radial magnetic energy during the nonlinear evolution of MRI mode (m=0, k=30) with
fixed mean flow. Three linear computations (at $t_1$--$t_3$) 
show no growth for k=30 mode. (e) Mean vertical magnetic field profiles for $t_1$--$t_3$ times 
during the nonlinear saturation used in the linear computations (m=0, k=30).}
\label{fig:fig3}
\end{figure}

We verify the stabilizing effect  
of the modified mean magnetic field
profiles by performing 
 linear computations of the saturated state, as described in section 3.  These computations use the \textit{background} 
profiles near the time of saturation. In this way we are able to 
calculate the linear growth rates of any resulting instabilities.
Each linear computation is initialized with perturbation of a single Fourier mode (m=0 and k=6) and
the equilibrium profiles. For each linear computation, the equilibrium profiles are the 
mean (m=0 and k=0 component) magnetic and velocity fields taken 
from the single mode nonlinear computation at a specific time near the saturated state (e. g. $\mathrm{t_1}$ 
marked in Fig.~\ref{fig:fig3}(a)).

We perform four separate linear computations for m=0, k=6 mode which 
are initialized with four different equilibrium profiles. Four set of 
equilibrium profiles are the mean fields 
taken from the single mode nonlinear computation at four times during the saturation. 
Figure~\ref{fig:fig3}(a) shows the radial magnetic 
energy vs. time for the nonlinear computation with fixed mean flow; the 
four times at which the linear computations are performed are also marked.
The linear growth rates calculated from the four linear computations 
decrease as the mode saturates from $\mathrm{t}_1$ to $\mathrm{t}_4$. 
The linear growth rate decreases from 0.58 (the value before the saturation 
at t=$\mathrm{t}_0$) to 0.048 at $\mathrm{t}_4$. 
The mean vertical magnetic field at these times is
also shown in Fig.~\ref{fig:fig3}(b).
 As the mean vertical magnetic field amplifies from $\mathrm{t}_1$ to $\mathrm{t}_4$ 
by the alpha effect 
in the plasma core, the resulting mode growth rate obtained from 
the linear computation substantially decreases (Fig.3 (c)).

We find that the dynamics of the saturation and 
the resulting alpha effect is independent of axial wave number k.  To illustrate this effect,  we have  performed nonlinear computation of a single mode with higher k (k=30)
shown in Fig.3 (d). The mode with k=30 grows faster than k=6 with 
a growth rate $\gamma \tau_{orbit} =1.17$
and starts to saturate around t=0.007. The saturation level for this mode is lower
than k=6 mode. Similarly, we initialize and 
perform linear computations with the saturated profiles from nonlinear 
computations at times $t_1$ -- $t_3$. The saturated mean vertical magnetic fields 
which are used as the equilibrium profiles for 
the linear computations are also shown in Fig. 3(e). 
Three separate linear computations are 
initialized at times $t_1$--$t_3$  shown in Fig. 3(d). 
Because the linear computations have been started 
with the saturated mean fields, 
 the linear growth rates obtained from these three computations 
are zero or negative, as seen in Fig.~3(d). 
The linear stability computations thus show that the amplification of 
the mean vertical magnetic
field by the azimuthal component of the 
alpha effect $<\widetilde V \times \widetilde B>_{\phi}$
 suppresses the mode growth and thus causes the mode saturation. The $\alpha$-effect of the saturated MRI therefore appears to be independent of the axial mode number $k$.

\subsection{Evolution with evolving mean flow}
\label{sec:nl2}
\begin{figure}
\includegraphics[]{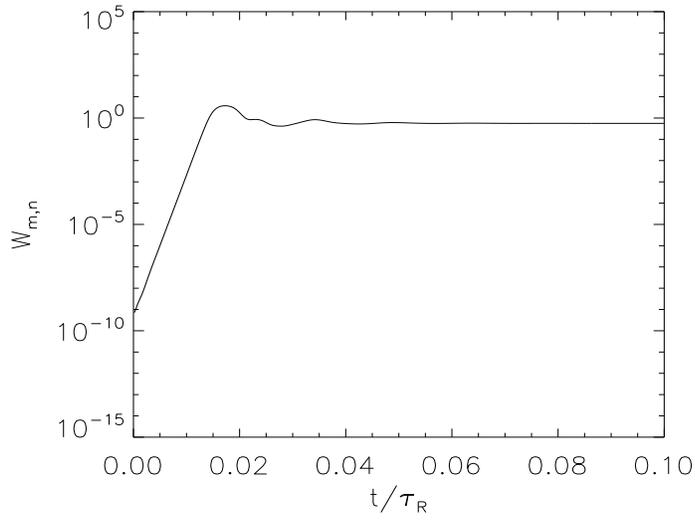}
\caption{Nonlinear evolution of axisymmetric MRI mode. 
Total magnetic energy $W_{m,n}=1/2 \int{\widetilde B_{(m,n}^2 dr^3}$ vs time for
computation when flow profile evolves ($\beta =10^4, P_m=1$, $S=10^2$).}
\label{fig:fig4}               
\end{figure}

\begin{figure}
\includegraphics[]{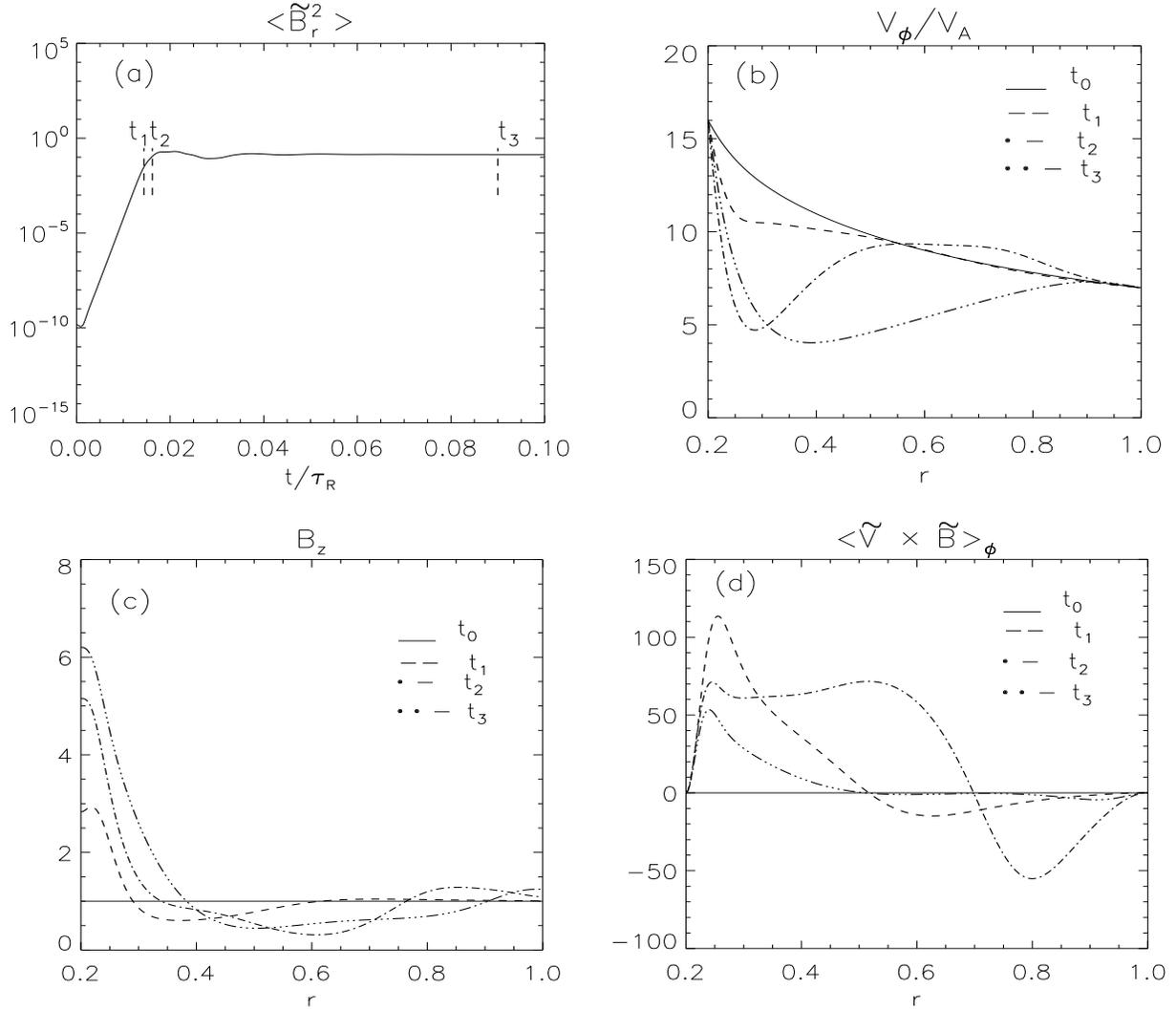}
\caption{Nonlinear m=0 mode computation with evolving flow. (a) averaged 
radial magnetic field fluctuations vs time. (b)-(d) radial profiles of
azimuthal flow, vertical magnetic field and the alpha effect 
(S $<\widetilde V \times \widetilde B>_{\phi}$) at 4 times.}
\label{fig:fig5}
\end{figure}

In the computation with a fixed mean azimuthal flow profile (Section 4.1), 
the mean magnetic field profile could be modified by the fluctuation-induced alpha effect. 
In the case when mean flow is allowed to evolve, the mean azimuthal flow profile will also be modified by the MHD stresses.
Figure \ref{fig:fig4} shows the nonlinear evolution of m=0, $k$=6 mode for 
evolving flow. Similar to the computation with fixed flow
(Fig.~\ref{fig:fig1}), axisymmetric mode grows and starts to saturate around t=0.016.
Figure \ref{fig:fig5} shows the evolution of magnetic field and
flow during the nonlinear computations at times $t=t_0$ (initial), $t_1,t_2$ 
( near saturation) and $t_3$ (final saturated). 
As is seen, the mean flow gradient is reduced in the core during the saturated state. 
The initial
 Keplerian mean flow profile flattens in the core (Fig. \ref{fig:fig5} (b)). Thus, with smaller mean flow gradient 
(less free energy for the mode growth), the mode starts to saturate. In addition to the mean flow modification,  
magnetic field generation still occurs. The magnetic field amplifies by a factor five in the core
(Fig.\ref{fig:fig5}(c)) as 
the azimuthal component of the alpha effect (Fig.\ref{fig:fig5}(d)) 
increases. Since the mean flow is allowed to evolve,
the amplification of the magnetic field is smaller than 
the computation with fixed mean flow. 

\begin{figure}
\includegraphics[]{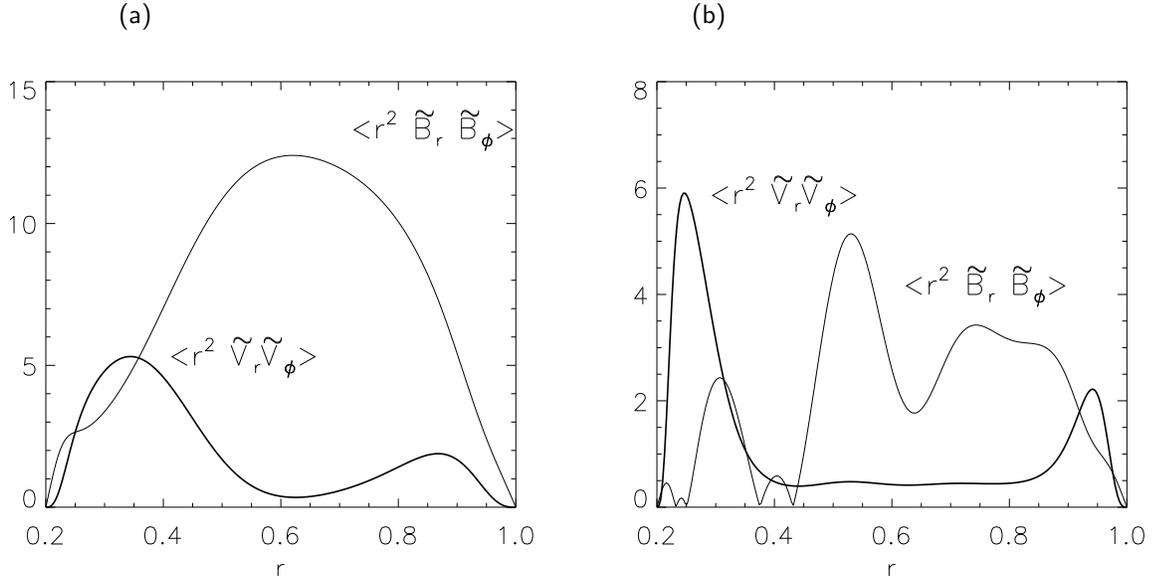}
\caption{Maxwell and Reynolds stresses profiles from nonlinear  
computation at t=$t_3$. (a) Pm=1 (b) Pm=0.1 }
\label{fig:fig6}
\end{figure}

\begin{figure}
\includegraphics[]{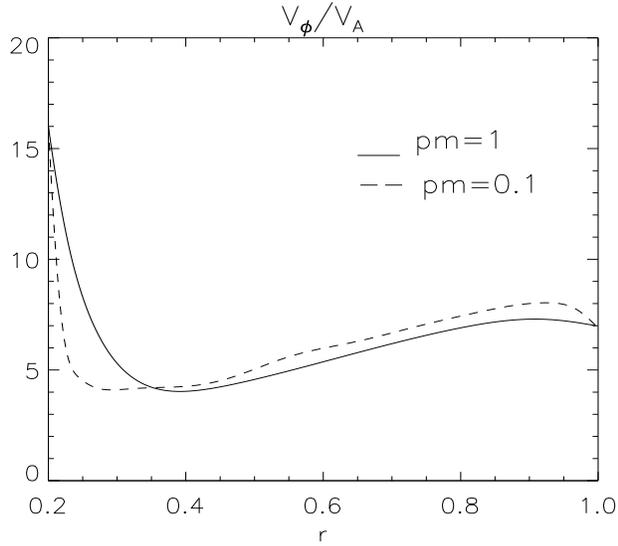} 
\caption{Final relaxed flow profiles from nonlinear computation 
for two Prandtl numbers, Pm=1 and Pm=0.1.}
\label{fig:fig7}
\end{figure}

The alteration in mean flow and mean magnetic field profiles both contribute to mode saturation.
The Reynolds and Maxwell stresses responsible for the momentum transport that flattens 
the flow profile are shown in Fig.~\ref{fig:fig6}(a).
Both stress terms contribute to the flow relaxation and outward momentum transport 
(consistent with Eqs.~\ref{eq:maxwell},~\ref{eq:reynold} in section 6).
 The ratio of the Maxwell and Reynolds stresses depends upon the magnetic
Prandtl number. Fig.~\ref{fig:fig6}(b) shows the profiles of Maxwell and Reynolds 
stresses for the  
same computations but with Pm=0.1 (final saturated state). 
It is seen that the Reynolds stress is larger than the Maxwell stress 
 in the core for computation with lower Prandtl number.
Final relaxed flow profiles for these computation with Pm=1 and Pm=0.1 is 
slightly different in the core (Fig.~\ref{fig:fig7}).

\section{Non-axisymmetric single mode nonlinear evolution}
\label{sec:asynl}
\begin{figure}
\includegraphics[]{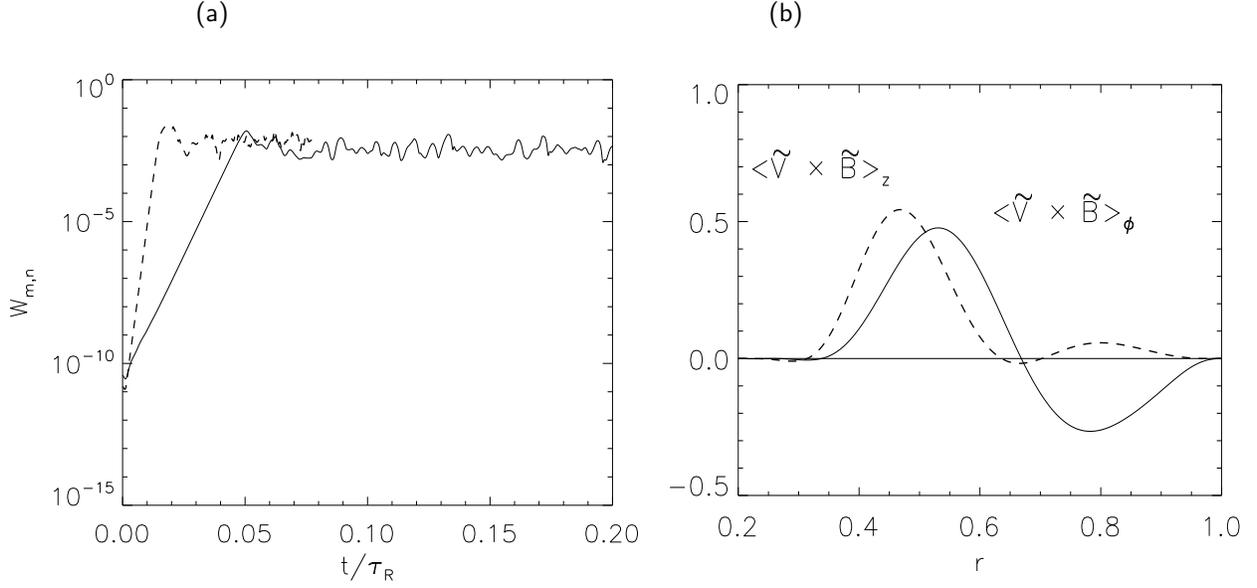}
\caption{Nonlinear evolution of non-axisymmetric m=1 MRI mode. 
(a) Magnetic energy $W_{m,n}=1/2 \int{\widetilde B_{r(m,n}^2 dr^3}$ vs time for
computation when mean flow profile is fixed for m=1 (solid line) and 
m=0 (dashed line) ($\beta =10^4, P_m=1$). (b) Azimuthal and vertical components of
the alpha effect during the linear growth of m=1 mode.}
\label{fig:fig8}
\end{figure}

\begin{figure}
\includegraphics[]{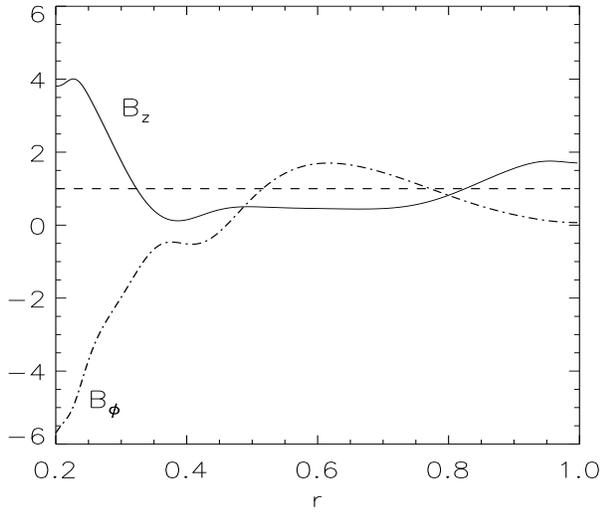}
\caption{Radial profiles of vertical and azimuthal magnetic field for m=1 mode 
at the time of saturation. Dashed line denotes mean magnetic profile ($B_z$) at t=0.}
\label{fig:fig9}
\end{figure}

Finally, we investigate the generation of magnetic field 
by the alpha effect from non-axisymmetric MRI modes. 
Nonlinear computation for an MRI mode with $m=1$ 
is performed with fixed Keplerian mean flow profile 
with the same equilibrium and parameters
as the $m=0$
mode. The non-axisymmetric m=1 mode grows more slowly than the $m=0$ mode, 
but saturates at the same level (Fig.~\ref{fig:fig8}). We find that 
the saturation mechanism for the m=1 mode is very similar to that of the m=0 mode. 
However for the $m=1$ mode, 
the vertical component of the alpha effect, $<\widetilde V \times \widetilde B>_{z}$,
 is also nonzero and becomes
 important in the process of mode saturation. The vertical component of the alpha effect  
will be calculated in the next section (Eq.~\ref{eq:vxbt},~\ref{eq:vxbz}) and shown to be nonzero 
for the non-axisymmetric MRI modes. Figure~\ref{fig:fig8}(b) shows the 
radial structure of both vertical and azimuthal components of the alpha effect
 for the m=1 mode during the linear growth.
Because of the nonzero vertical alpha effect, an azimuthal magnetic field $B_{\phi}$ 
(which is zero at t=0) is generated and is shown in Fig.~\ref{fig:fig9}. 
Thus, the saturation of m=1 mode occurs through 
amplification of both vertical and azimuthal magnetic field through the alpha effect.

\section{Quasilinear calculations}
\label{sec:ql}
To evaluate the influence of the instability (fluctuations) on the mean magnetic 
field, we calculate the alpha effect --- the mean 
fluctuation-induced electromotive force $<\widetilde V \times \widetilde B>$,  
where $\widetilde V$ and $\widetilde B$ are
the velocity and magnetic fluctuations, and $<>$ denotes an average over 
the azimuthal and vertical directions.
Perturbed quantities are assumed in the form of
$exp(\gamma t + im\phi+ik z)$. The alpha effect is obtained for a single MRI 
Fourier mode (single m and $k$). The linearized momentum equations in terms of the Lagrangian 
displacement $\mathbf{\xi}$ are (Chandrasekhar 1961),
   
\begin{eqnarray}
 (\bar{\gamma}^2 + \omega_A^2 + 2 r \Omega \Omega^{'}) \xi_r - 2\bar{\gamma} \Omega \xi_{\phi} = - \frac{\partial X}{\partial r}\label{eq:mri1}\\
(\bar{\gamma}^2 + \omega_A^2) \xi_{\phi} + 2\bar{\gamma} \Omega \xi_{r} = -\frac{im X }{r}\\
\label{eq:mri2}
(\bar{\gamma}^2 + \omega_A^2) \xi_{z} = -i k X
\label{eq:mri3}
\end{eqnarray}
 where $\bar{\gamma} = \gamma + i m \Omega$, $X = \widetilde{P} + B_{0} \cdot \widetilde{B}$, 
$\omega_A = k^2 B_{0}^2/\rho$ and $\Omega = V_{\phi}/r$ is angular velocity. 
The velocity fluctuations are related to the Lagrangian 
displacement $\xi$,  $\widetilde V_r = \bar{\gamma} \xi_r $, 
$\widetilde V_{\phi} = \bar{\gamma} \xi_{\phi}- [\frac{\partial V_{\phi}}{\partial r}-\frac{V_{\phi}}{r}] \xi_r $ and $\widetilde V_z = \bar{\gamma} \xi_z $. 
In an ideal plasma (zero resistivity) magnetic field
perturbations are also directly related to the displacement, 
$\widetilde{\mathbf{B}} = i k B_0 \mathbf{\xi}$.
Using incompressibility ($\xi_z = -1/ikr (r \xi_r)^{\prime} -m/rk \xi_{\phi}$) 
and Eqs.~\ref{eq:mri1}-\ref{eq:mri3}, azimuthal and vertical displacements are obtained in terms of
$\xi_r$,  
\begin{equation} 
\xi_{\phi} = \frac{1}{(1+m^2/(r^2 k^2)}[\frac{-2 \Omega \bar{\gamma}}{(\bar{\gamma}^2+\omega_A^2)} \xi_r+ \frac{i m}{r^2 k^2}(r \xi_r)^{\prime}]
\label{eq:xiphi}
\end{equation} 
\begin{equation}
\xi_{z} = \frac{-mkr}{(r^2 k^2+m^2)}[\frac{-2 \Omega \bar{\gamma}}{(\bar{\gamma}^2+\omega_A^2)} \xi_r+ \frac{i }{r k}( 1- \frac{m^2}{(r^2 k^2+m^2)})(r \xi_r)^{\prime}]
\label{eq:xiphi}
\end{equation} 
We then construct the quasilinear alpha effect  
, $<\widetilde V \times \widetilde B>_{z}=1/2 Re(\widetilde V_r^{*} \widetilde{B}_{\phi} - 
\widetilde V_{\phi}^{*} \widetilde{B}_r) $ and $<\widetilde V \times \widetilde B>_{\phi} = 1/2 Re(\widetilde V_z^{*} \widetilde{B}_r - 
\widetilde V_r^{*} \widetilde{B}_z)$, in terms of the radial displacement $\xi_r$,  
\begin{equation}
<\widetilde V \times \widetilde B>_{\phi} = \frac{\gamma B}{r} \xi_r (r \xi_r)^{\prime} -  \frac{ m^2 \gamma B}{r (m^2 + k^2 r^2)} \xi_r (r \xi_r)^{\prime} + \frac{2m^2 \gamma B \Omega^2}{r H^2}(\omega_A^2 - \gamma^2) \xi_r^2 - \frac{2m^4 \gamma B \Omega^4}{r H^2} \xi_r^2 
\label{eq:vxbt}
\end{equation}

\begin{equation}
<\widetilde V \times \widetilde B>_{z} = -  \frac{ m \gamma k B}{(m^2 + k^2 r^2)} \xi_r (r \xi_r)^{\prime} + \frac{2m \gamma k B \Omega^2}{H^2}(\omega_A^2 - \gamma^2) \xi_r^2 - \frac{2m^3 \gamma k B \Omega^4}{ H^2} \xi_r^2 
\label{eq:vxbz}
\end{equation}
where,
\begin{equation}
H^2 = (1+\frac{m^2}{k^2 r^2})(4 m^2 \gamma^2 \Omega^2 + (\gamma^2 -m^2 \Omega^2 +\omega_A^2)^2) 
\end{equation} 
For axisymmetric m=0 modes, the alpha effect reduces to a simple form. The vertical
component of the alpha effect (Eq.~\ref{eq:vxbz}) vanishes. The azimuthal component
 (Eq.~\ref{eq:vxbt}) is nonzero and becomes,\\
\begin{equation}
<\widetilde V \times \widetilde B>_{\phi} = \frac{B_0}{\gamma r} \widetilde{V}_r \frac{\partial}{\partial r}(r \widetilde{V}_r)
\label{eq:dynamo}
\end{equation} 
Thus, the instability is accompanied by an alpha effect that generates magnetic field.

\begin{figure}
\includegraphics[]{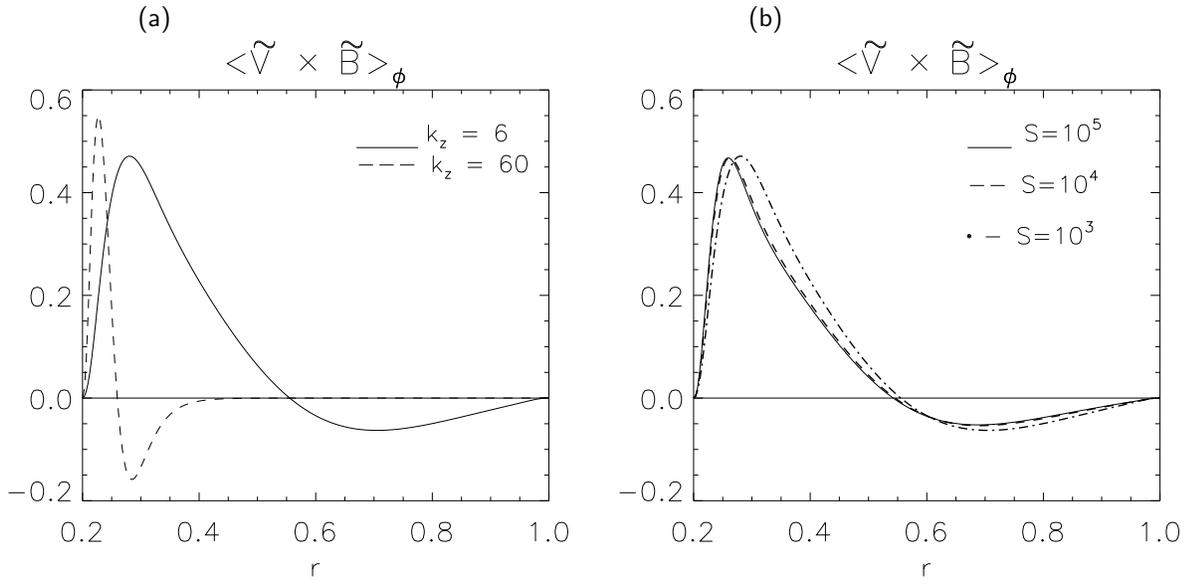} 
\caption{Radial structure of the alpha effect $<\widetilde V \times \widetilde B>_{\phi}$ for 
the m=0 MR mode (a) 
for two wave numbers $\mathrm{k =6, 60}$ ($S=10^3$), 
 (b) for $S=10^3,10^4,10^5 $ with $k =6$.
 $V_{\phi}/V_A = 51$, $\beta=10^5, Pm=1$. The growth rates 
are $\gamma \tau_{orbit} =0.23$ and $\gamma \tau_{orbit} =1.7$ 
for $k=6$ and $k=60$ respectively ($\tau_{orbit}=2 \pi/\Omega$).}
\label{fig:fig10}
\end{figure}

To obtain the radial structure of the alpha effect, linear computation
(Eqs. 1-6) is performed for axisymmetric modes. 
We construct the alpha effect using linear eigenfunctions obtained from 
single mode computation. Figure \ref{fig:fig10}(a) 
 (solid line) shows the azimuthal component of the alpha effect for the m=0 mode with $k =6$. 
The azimuthal component generates axial magnetic field in the inner region of the disk,
 a stabilizing influence. At higher k values (Fig. \ref{fig:fig10}(a), dashed line)
the mode structure and alpha effect becomes more localized radially.
Linear computations 
are also performed for different Lundquist numbers, $S=10^3,10^4,10^5$ (Fig. \ref{fig:fig10}(b)) . 
The growth rate of the MRI mode does not change significantly with Lundquist number 
(since MRI is an ideal mode). 
It is seen that the structure is 
 slightly broader for lower S, which is expected 
because of higher dissipation at lower S. 
Similar radial mode structure ($\widetilde V_r$) with peaked structure near the inner radius
has also been obtained in earlier work by Curry, Pudritz \& Sutherland 1994 and Blokland et al. 2005.
 
We can similarly evaluate the fluid stresses that alter the flow. 
The MHD stresses for axisymmetric MRI mode are given by Balbus and Hawley 2002.
Here we also calculate the quasilinear stresses for a non-axisymmetric MRI mode.
The nonlinear evolution of this mode was examined in section \ref{sec:asynl}. 
Using eqs~\ref{eq:mri1}-\ref{eq:mri3},  
the Maxwell stresses (for both
axisymmetric and non-axisymmetric modes) are calculated 
in terms of radial displacement
\begin{equation}
< r^2 \widetilde B_r \widetilde B_{\phi}> = -\frac{r^2 k^2 B^2 \gamma \Omega (\gamma^2 +\omega_A^2)}{H^2} \xi_r^2 - \frac{m^2 r^2 k^2 B^2 \gamma \Omega^2}{H^2} \xi_r^2  
\end{equation}
\begin{equation}
< r \widetilde B_r \widetilde B_{z}> = \frac{m k B^2 \gamma \Omega (\gamma^2 +\omega_A^2)}{H^2} \xi_r^2 + \frac{m^3 k B^2 \gamma \Omega^3}{H^2} \xi_r^2  
\end{equation}
It is seen that the vertical Maxwell stress is zero for the axisymmetric 
modes and the azimuthal Lorentz force 
$< \widetilde J \times \widetilde B>_{\phi} = \frac{1}{2r^2}\frac{d}{dr}<r^2 B_r B_{\phi}>$ is obtained
\begin{equation}
 < \widetilde J \times \widetilde B>_{\phi}={\frac{ k^2}{ r^2 }\frac{d}{d r} (D \Omega)}
\label{eq:maxwell}
\end{equation}
\begin{equation}
{D=\frac{r^2 B^2 |\widetilde V_r|^2}{\gamma (\gamma^2 +\omega_A^2)}}
\label{eq:diff}
\end{equation}
Similarly, 
the Reynolds stresses are, 
\begin{equation}
< r^2 \widetilde V_r \widetilde V_{\phi}> = -\frac{r^2 \gamma^3 \Omega (\gamma^2 +\omega_A^2)}{H^2} \xi_r^2 - \gamma r^3 \Omega^{\prime} \xi_r^2 - \frac{m^2 r^2 \gamma \Omega^3}{H^2} \xi_r^2  - \frac{m^4 r^2 \gamma \Omega^5}{H^2} \xi_r^2
\end{equation}
\begin{equation}
< r \widetilde V_r \widetilde V_{z}> = \frac{m \gamma^3 \Omega (\gamma^2 +\omega_A^2)}{k H^2} \xi_r^2 + \frac{m^3 \gamma \Omega^3 (2 \gamma^2 +\omega_A^2)}{k H^2} \xi_r^2 + \frac{m^5 \gamma \Omega^5}{k H^2} \xi_r^2 
\end{equation}
 Again, the only nonzero component for m=0 modes is azimuthal,  
 and $< \widetilde V \cdot \nabla \widetilde V>_{\phi} = \frac{1}{2r^2}\frac{d}{dr}<r^2 V_r V_{\phi}>$ is calculated to be,

\begin{equation}
 <\widetilde V \cdot \nabla \widetilde V>_{\phi}={\frac{ \gamma^2}{ (r^2 B^2) }\frac{d}{d r} (D \Omega)} + {\frac{ 1}{ (2 r^2 \gamma) }\frac{d}{d r} (r^3 \Omega^{\prime}|\widetilde V_r|^2)}
\label{eq:reynold}
\end{equation}
where the coefficient D is defined in Eq.~\ref{eq:diff}.
Adding eqns~\ref{eq:maxwell} and ~\ref{eq:reynold}, the total azimuthal fluid
force for m=0 modes becomes,
\begin{equation}
 < \widetilde J \times \widetilde B>_{\phi}+ \rho <\widetilde V \cdot \nabla \widetilde V>_{\phi}=(k^2 + \frac{\gamma^2}{B^2})\frac{d}{r^2 d r} (D \Omega)+ {\frac{ 1}{ (2 r^2 \gamma) }\frac{d}{d r} (r^3 \Omega^{\prime}|\widetilde V_r|^2)} 
\label{eq:totalforce} 
\end{equation}
The fluid force consists of a simple diffusion-like part (the first term) and a more complex part 
(the second term) that also depends on flow nonuniformity ($\Omega^{\prime}$).

\section{Summary}
\label{sec:summ}
In summary, we have shown that the MRI will generate large-scale 
magnetic field as a mechanism for saturation. Computation of the fully evolved, nonlinear steady state of a single mode establishes 
that the mechanism that determines mode saturation is as described by quasilinear theory, 
and that magnetic field generation persists. 
This is an 
alpha effect, and can be understood through the stabilizing 
effect of the magnetic field.  The analytic and computational 
work presented here treats one mode, as a means to clarify 
the effect.  The influence of multiple modes and mode coupling 
will be of considerable importance. In a future work we will investigated the MRI saturation 
via the alpha effect using the full nonlinear computations with multiple modes. The field generated is 
large-scale in that it is symmetric in toroidal and axial directions. 
However, it varies radially such that the total toroidal and 
axial magnetic flux within the plasma does not change 
significantly in time. Thus, the examples studied here 
would not be defined as a ``large-scale dynamo.''
 Future work also will investigate whether there are 
geometric variations for which the generation of 
large-scale field would be accompanied by the generation of total magnetic flux.\\

\acknowledgments
We acknowledge useful discussions with Ellen Zweibel. We also thank the referee for useful comments.
This work is supported by National Science Foundation. 


\end{document}